\documentclass[12pt,preprint]{aastex}

\shorttitle{High Spatial Resolution Spectroscopy of W51 IRS2E and IRS2W}
\shortauthors{Barbosa et al.}

\begin{document}


\title{High Spatial Resolution Spectroscopy of W51 IRS2E and IRS2W:
Two Very Massive Young Stars in Early Formation Stages}


\author{C\'assio L. Barbosa\altaffilmark{1}}
\affil{IP\&D -- UNIVAP, Av. Shishima Hifumi, 2911, 12244-000 S\~ao Jos\'e dos
Campos, SP,  Brazil}
\email{cassio@univap.br}

\author{Robert D. Blum\altaffilmark{2}}
\affil{NOAO Gemini Science Center, 950 Cherry Ave., Tucson, AZ 85719, USA}
\email{rblum@noao.edu}

\author{Peter S. Conti\altaffilmark{3}}
\affil{JILA, University of Colorado, Boulder, CO 80309-0440, USA}
\email{pconti@jila.colorado.edu}

\author{Augusto Damineli\altaffilmark{4}}
\affil{IAG -- USP. Rua do Mat\~ao, 1226, 05508-900, S\~ao Paulo, SP, Brazil}
\email{damineli@astro.iag.usp.br}
\and

\author{Elysandra Figuer\^edo\altaffilmark{5}}
\affil{The Open University - Walton Hall, Milton Keynes MK7 6AA, UK}
\email{e.figueredo@open.ac.uk}



\begin{abstract}
We present $K$-band spectra of the near infrared counterparts to IRS2E
and IRS2W which is associated with the ultracompact HII region W51d, both of
them embedded sources in the Galactic compact HII region W51 IRS2. The high
spatial resolution observations were obtained with the laser guide star
facility and Near infrared Integral Field Spectrograph (NIFS) mounted at the
Gemini North observatory. The spectrum of the ionizing source of W51d shows
the photospheric features NIII (21155 \AA) in emission and HeII (21897
\AA) in absorption which lead us to classify it as an young O3 type
star. We detected CO overtone in emission at 23000 \AA \ in the
spectrum of IRS2E, suggesting that it is a massive young object still
surrounded by an accretion disc, probably transitioning from the hot
core phase to an ultracompact HII region.
\end{abstract}


\keywords{stars: early type --- stars: formation --- infrared: stars ---
techniques: spectroscopy}



\section{Introduction}

The mechanisms of massive star formation ($M_{*}>10 M_{\sun}$) are the
subject of intense discussion even after two decades of great
progress and improvement of observational techniques. In the last two
decades, the development of larger and more sensitive infrared
detectors associated with the commissioning of 8-m class telescopes
equipped with adaptive optics has made it possible to observe ever
deeper into the sites of massive star formation at relatively short
wavelengths. Yet, while all these improvements
have contributed to our understanding of massive star formation, the
basic process, i.e., formation via an accretion disk, coalescence,
competitive accretion, or a combination of these is still unknown
\citep[for a review on this discussion see][]{zin07}. The {\it Spitzer}
Space Telescope is providing an unprecedented view at longer wavelengths
and soon the $Herschel$ Space Observatory and ALMA will combine very high
sensitivity and angular resolution to probe the earliest phases of massive star
formation.

W51 is one of the most luminous complexes of massive star forming
regions in the Galaxy \citep{gold94} with multiple HII regions
\citep{wilson70} that host at least six HII regions with embedded
clusters, all of them optically obscured \citep{kumar04}. The most
luminous of all HII regions is G49.5-0.4 which can be resolved into
mainly two bright infrared sources: IRS1 and IRS2, which are coincident with
radio sources W51e and W51d respectively \citep{mar72}. IRS2 is a
compact HII region harboring three ultracompact HII (UCHII) regions
W51d, W51d$_{1}$ and W51d$_{2}$ \citep{mehr94} and a plethora of
infrared sources \citep{gold94,oku01, oka01, kram01, fig08, barb08}. The most
prominent sources in the $K$
band are named IRS2 East and IRS2 West (or IRS2E and IRS2W,
respectively, for short). IRS2W was identified as a NIR and MIR peak
of nebular emission \citep{gold94} and \citet{oka01} have correlated it with the
UCHII region W51d.

In this letter we report the results of spectroscopic observations taken with
laser guide star (LGS) adaptive optics of IRS2E and the UCHII region W51d
(IRS2W). In the section \ref{obs} we report the technical details of
the observations and in the section \ref{res} we report and discuss the results.
Finally, in the section \ref{conc} we summarize our conclusions.

\section{Observations\label{obs}}

The data were obtained with the Near infrared Integral Field
Spectrograph (NIFS) mounted on the Frederick C. Gillett 8--m telescope
at the Gemini North observatory on Mauna Kea, HI, USA, in queue mode
on May, 4th 2007 under program ID GN-2007A-Q-34. NIFS provides 2D
integral field spectroscopy with R$\sim$5200 over a 3\arcsec$\times$3\arcsec\ 
field of view with 0.043$\times$0.10'' rectangular ``pixels''. Also,
we used the Gemini North Adaptive Optics system Altair with a LGS to
achieve an angular resolution of $<$0.2\arcsec\ measured upon the 2D images
build from the datacubes.

The observations were carried out with the standard procedures for the
near infrared (NIR): we observed IRS2 in a series of three pointings of 1000
seconds on source and 1000 seconds on a ``blank" sky position at $\sim10\arcsec$
north of IRS2 for each field. The latter image was used to subtract the sky
emission from the images of the targets. We also observed the A0V star HIP 98640
with similar airmass as ISRS2 in order to remove the telluric absorption lines
from the spectra of the observed targets. The spectra of A0V stars show only
the Br$\gamma$ feature in absorption in the $K$ band.  We
eliminated this feature by fitting a Voight profile to the Br$\gamma$
leaving a spectrum with only telluric absorption features. This
telluric spectrum was used to correct the spectra of the targets.

Wavelength calibration was achieved with an exposure of an Argon arc
lamp obtained with the Gemini calibration module GCAL. All data were
precessed under IRAF\footnote{IRAF is distributed by the National
Optical Astronomy Observatories, which are operated by the
Association of Universities for Research in Astronomy, Inc., under
cooperative agreement with the National Science Foundation.}
environment with scripts written for NIFS available through the Gemini IRAF
package. We extracted the spectra through a 0.4\arcsec\ circular
aperture centered on the sources. This aperture was chosen to minimize
the contribution of the nebular emission from the HII region.

\begin{figure}
\includegraphics[angle=270,scale=.70]{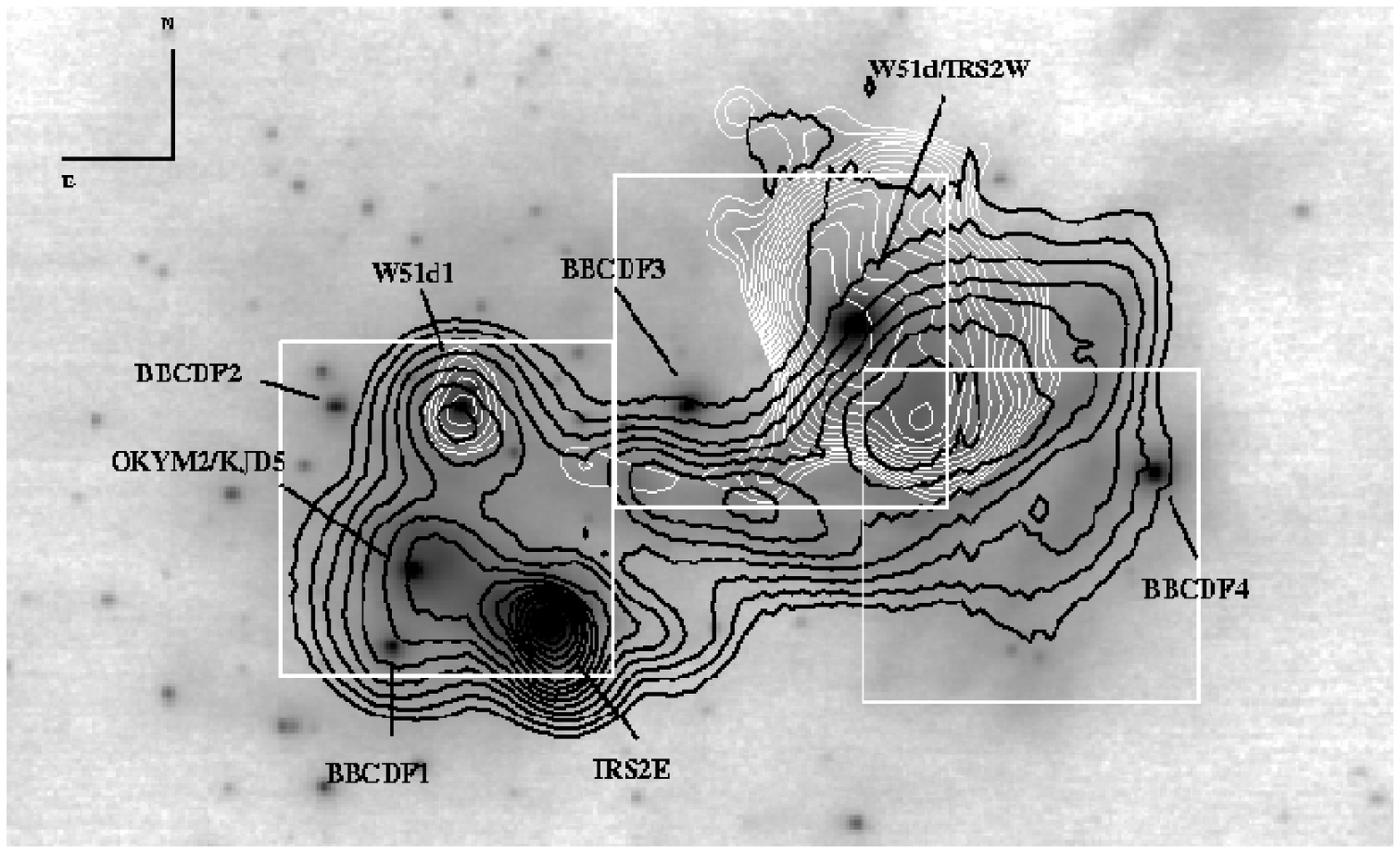}
\caption{Finding chart of IRS2. The white contours are the radio continuum
emission at 2 cm. The black contours represent the MIR emission at 12 \micron.
White squares represent the field of view of NIFS (3\arcsec$\times$3\arcsec)
and  indicate the positions of the three pointings of our program. The sources
labeled are those which we have ``pseudo long-slit" spectra. All contours are in
arbitrary units.\label{find_nifs}}
\end{figure}

We show in Figure \ref{find_nifs} the finding chart of IRS2. It is composed by
the high resolution image taken with the adaptive optics NIR camera NACO in
the $K$ band. This image is available at the ESO public database. We overplotted
the radio continuum emission at 2 cm from \cite{wc89} in white contours. The
black contours represent the MIR emission at 12 \micron\ from the Gemini South
MIR camera T-ReCS. All NIFS pointings are represented by white squares and the
sources which we have spectra are labeled.
In the following section we present the results of the observations of IRS2E and
the UCHII region W51d in the ``pseudo long-slit" mode extracted from the
images delivered by NIFS. A complete analysis of these data, including the
spectra of all observed sources and the results from the IFU mode, such as the
velocity maps of the regions observed will be presented in a forthcoming paper
\citep{barb08}.

\section{Results and Discussion\label{res}}

\subsection{IRS2E}

\begin{figure}
\includegraphics[angle=270,scale=.70]{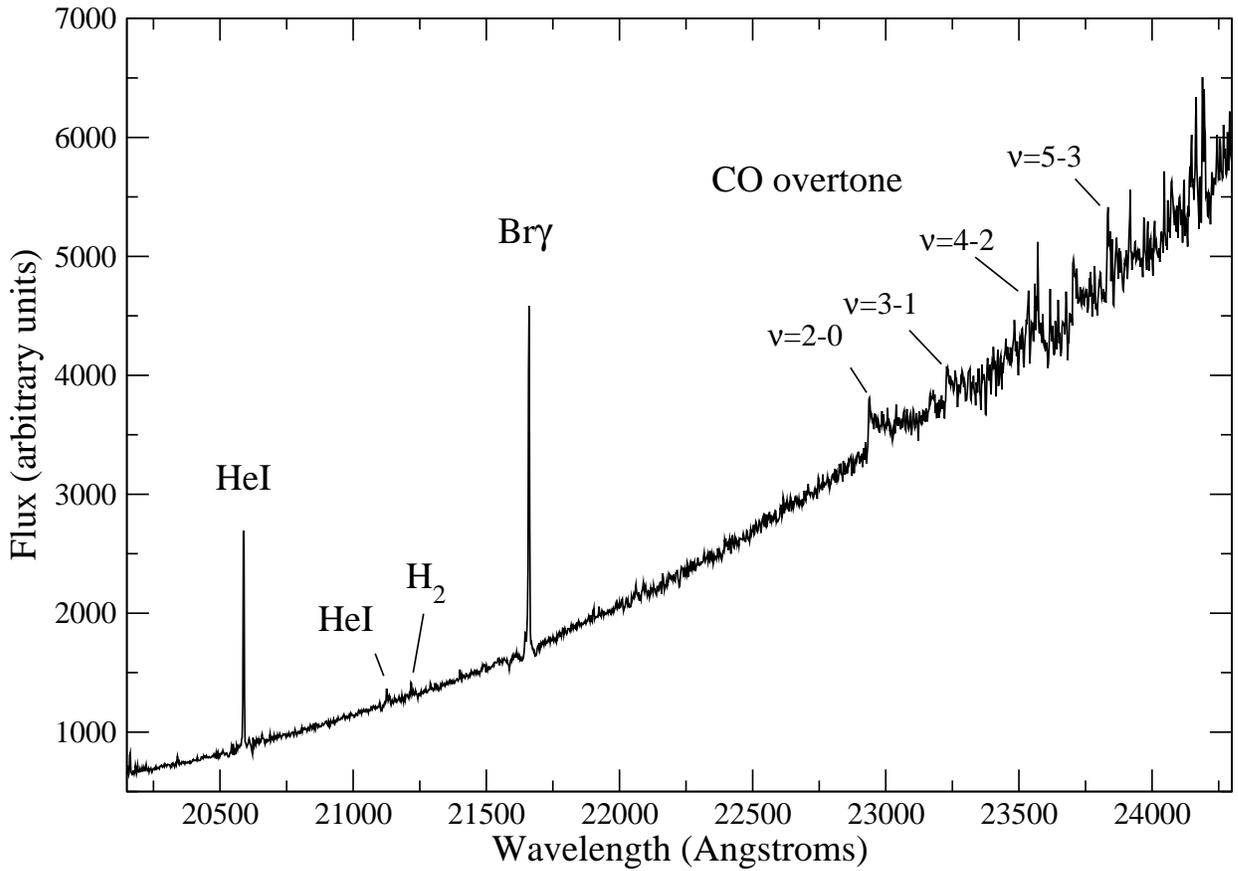}
\caption{$K$-band spectrum of source IRS2E. After the spectrum was extracted it
was multiplied by a black body of 9000 K to recover the true spectral
shape of the source. The flux is in arbitrary units.\label{irs2e}}
\end{figure}

Figure \ref{irs2e} shows the spectrum of IRS2E, the brightest $K-$band source
in the compact HII region. It was obtained from the first pointing of the
telescope. The continuum shows a steep, rising slope to longer
wavelengths reflecting the fact that this source is under a heavy extinction:
A$_{V}\sim$63 mag according to our unpublished MIR data. This high extinction
makes IRS2E undetected at wavelengths shorter than 16000 \AA\footnote{
Unpublished images obtained from the AO assisted camera NACO at the VLT shows
that the source detected at the $J$ band by previous authors is just a bright
knot of gas.}. The spectrum of
Figure \ref{irs2e} shows strong lines of HeI (20587 \AA) and unresolved
Br$\gamma$ (21661 \AA) that may arise from the circumstellar gas heated by hot
stars. We also note nebular lines of HeI (21127 \AA) and the H$_{2}$ (1--0 (S1),
21128 \AA) that come from the diffuse nebular emission of the HII region.
Previous authors \citep[e.g.][]{oku01,fig08} describe the presence of [FeIII]
lines in the spectrum of IRS2E. These emission lines are typical of shocked gas
in the vicinity of MYSOs \citep{han02,bik06}. The spectrum in the Figure
\ref{irs2e} does not show any of these lines. This is due to the small aperture
used to extract the spectrum. Actually, our unpublished line maps show that
[FeIII] emission comes from the extended region between IRS2E and OKYM2/KJD5.

More interesting, IRS2E shows CO overtone emission as we detected the (2--0),
(3--1), (4--2) and (5--3) bandheads between 22935 \AA\ and 23838 \AA. Previous
NIR spectroscopy of IRS2E \citep[e.g.,][]{oku01,fig08} could not detect the CO
bandhead emission due to the large contribution of nebular emission through
wide and long slits.

The detection of the CO overtone in emission is associated with the
presence of an accretion disc, both in low-mass \citep[e.g.,][]{naj96}
and high-mass YSOs \citep{bik04,blum04}. The CO overtone emission comes from the
inner parts a circumstellar disk, in regions of high column densities
($10^{20}$--$10^{21}$ cm$^{2}$) and temperatures between 1500--4500 K.

IRS2E is a deeply embedded
source which is bright at wavelengths longer than 20000 \AA\ and is
not associated with any UCHII region. These characteristics are those
expected for a very young and massive YSO that is transitioning to an UCHII
region; the high accretion rate presumably prevents the formation of an UCHII
region as the ionized gas near the source falls onto the star, according to the
evolutionary scenario proposed by \citet{church02}.  This class of
object was observed in NGC 3576 by \citet{barb03}, for example.

\subsection{W51d}

\begin{figure}
\plotone{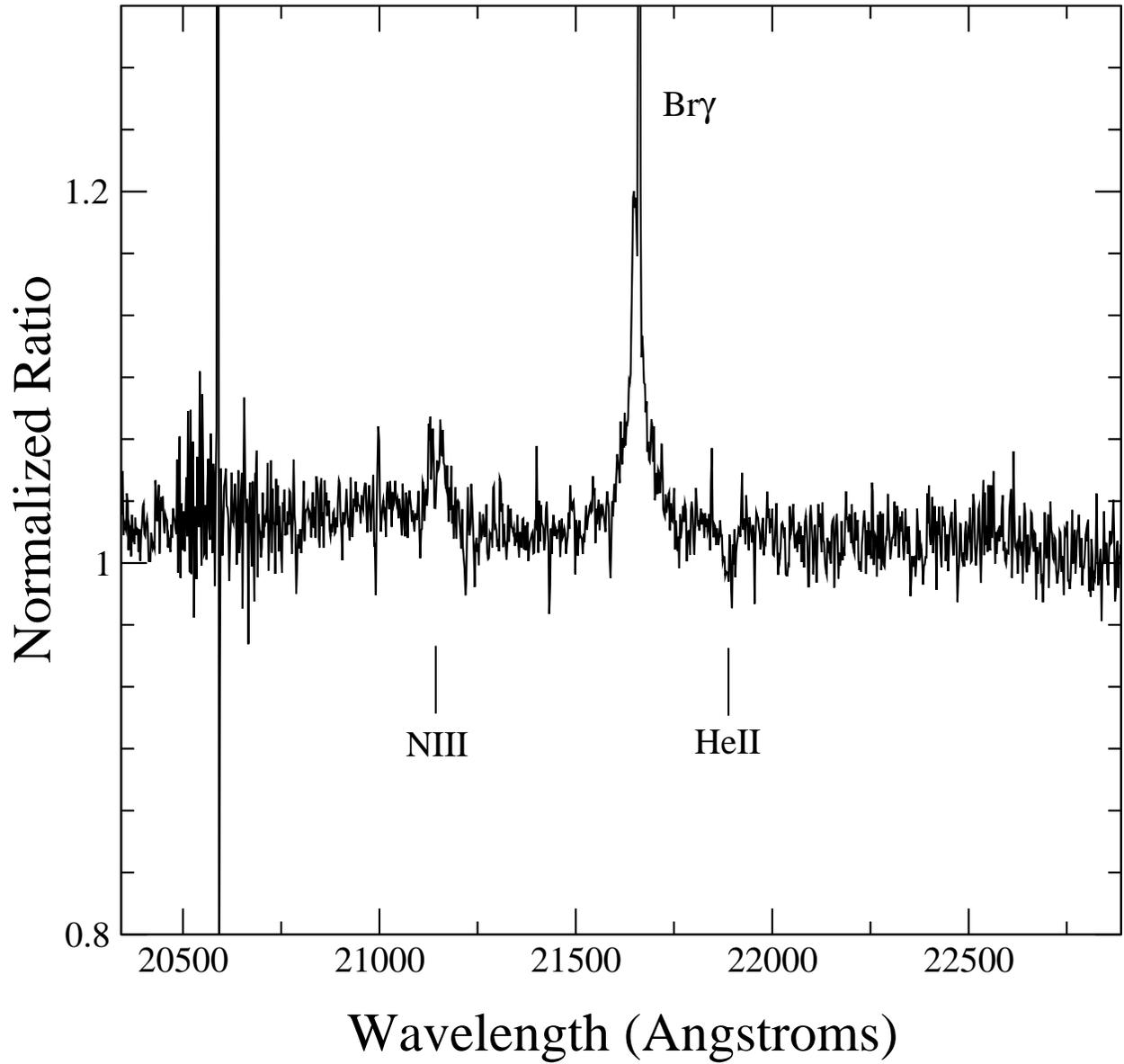}
\caption{$K$-band spectrum of source W51d. This spectrum was zoomed to show the
NIII multiplet and HeII. The thin line near 20587 \AA\  is a residual feature
produced after the correction of the telluric lines.\label{w51d}}
\end{figure}

Figure \ref{w51d} shows the spectrum of the UCHII region W51d. Treated as a peak
of nebulosity in the NIR by \citet{gold94}, \cite{fig08}
identified a stellar object as responsible for the ionization of this
region based on the correlation between the radio and MIR emission presented by
\citet{oka01} and the new high resolution images obtained by us or collected at
public databases.

The spectrum of W51d was normalized and zoomed near the intense Br$\gamma$ line
to better show the detected features. This line has a broad component ($\sim$
700 km s$^{-1}$), probably originated in fast winds coming from the hot ionizing
star and two scenarios are possible to explain it. In the first case, the very
young star has strong winds that blew away its surrounding dust. In the second
case, the ionizing star is still surrounded by a torus that is seen more face
on. However, only higher resolution data could favor one of
these scenarios. Besides this nebular line, we detected photospheric lines
corresponding to the NIII multiplet near 21155 \AA\ in emission and HeII at
21897 \AA\ in absorption.

The ionizing source of W51d can be classified as an early O star,
given the detected features in the spectrum of Figure
\ref{w51d}. Comparing this spectrum with the catalog of spectra of
optically visible OB stars presented by \citet{han05} we can constrain
its spectrum type to an O3--4 type star. The spectrum of an optically
visible O3V star shows both the NIII multiplet (in emission) and the
HeII (in absorption) at wavelengths longer than 20000 \AA. Although
our spectrum cannot establish its spectral subtype, the absence of CIV
(20780 \AA) in emission, as seen in mid type O stars, favors the
identification of this object as a star hotter and more massive than
type $\sim$ O4. It would make this object one of the most massive
ionizing source of an UCHII region \citep{wh97,han02,bik05,blum08}.

\citet{wc89} and \citet{mehr94} classify the ionizing source of W51d as an O5.5
and O5 respectively, both based on radio continuum emission assuming a distance
of 7 kpc to this source. The disagreement
between the results based on radio data and the spectroscopy
can be explained if the massive YSO had enough time to clear the circumstellar
material, lowering its density (and the extinction as well). In this case, the
star will not be as embedded as a hot core (like IRS2E) and some fraction of
the total number of the Lyman continuum photons emitted by the star will freely
escape without ionizing the surrounding gas. If this is the case, the ionizing
source would be of earlier spectral type.

\subsection{The Distance to IRS2}

Normally, one could derive the distance to W51d using its photometry in the NIR
and the spectral type obtained above. However, we could not obtain a reliable
photometry in the $H$ band from the NACO image. W51d seems to be saturated (as
well as the object BBCDF4) and every attempt to reconstruct its PSF has failed.
Moreover, this source seems to have strong excess emission and in this case we
cannot derive its $H-K$ color nor assign its extinction. Without its $H-K$ color
and extinction we cannot derive an estimate for its distance using
spectroscopic parallax.

On the other hand, based on the spectral type presented in the previous
section, we can estimate the distance to IRS2 using the radio data available in
the literature. We can derive the number of photons emitted in the Lyman
continuum by an O3 star and scale the distance until we reproduce the
flux measured in the radio at a given wavelength. We need to keep in
mind the limitations of this method, since it assumes that all photons
emitted in the Lyman continuum are effectively ionizing the gas, and
this may not be true as explained in the previous section.  Some
amount of the total number of photons may be escaping the UCHII while
other photons are destroyed by dust.

Starting with equation (4) given by \cite{kcw94} ($\xi=1$, i.e. all photons in
the Lyman continuum effectively ionize the gas) we obtain the distance $d$ (in
kpc) as a factor of the electron temperature $T_{e}$,
the correction factor $\alpha(\nu,T_{e})$, the frequency $\nu$ and the flux
density $S_{\nu}$ as:

\begin{equation}
d^{2}=1.32\times10^{-49}N_{C}\frac{\alpha(\nu,T_{e})}{S_{\nu}(Jy)}\Big(
\frac{\nu}{GHz}\Big)^{-0.1}\Big(\frac{T_{e}}{K}\Big)^{0.5}.
\end{equation}

The critical parameter in the above equation is the number of photons emitted in
the Lyman continuum $N_{C}$. Values for that parameter can vary an order of
magnitude depending on which author is quoted. We derived $N_{C}$ as follow: for
spectral types ranging from an O3 to an O4 star at the zero age main sequence
(ZAMS) we took their effective temperature from Table 1 of \cite{mart05}:
44,616 K for an O3 and 43,419 K for an O4. Using the Geneva models of
\cite{scha92} we obtained the bolometric magnitude $M_{bol}$ for each star in
the ZAMS (-8.94, O3 and -8.7, O4). We derived the stellar radius from the
effective temperature above and its corresponding luminosity: 5.476 and 9.2
$R_{\sun}$ for an O3 and 4.638 and 8.7 $R_{\sun}$ for an O4. The radius and
the luminosity calculated for a star in the ZAMS are significantly smaller than
those quoted for main sequence stars and may be more reliable for W51d, given
its young age. Finally, we used the radius for each star to obtain the number
of ionizing photons emitted per unit time from the equation (9) and the
ionizing fluxes $Q_{0}$ for each respective effective temperature also quoted
in the Table 1 of \cite{mart05}. According to these procedures an O3 can
produce $1.86\times10^{49}$ s$^{-1}$ and an O4, $1.44\times10^{49}$ s$^{-1}$ in
the ZAMS.

Returning to our equation (1), $S_{\nu}$ and $T_{e}$ quoted for W51d are:
5.35 Jy and $10^{4}$K at $\nu=$15 GHz \citep{wc89} and  $\alpha(\nu,T_{e})=$
0.9767 \cite{mez67}. The radio flux of W51d is blended with that of W51d$_{1}$
in data presented by \cite{mehr94}, so it is overestimated and therefore cannot
be used. The results are: for a ZAMS O3 star $d=5.8$ kpc and for an ZAMS O4
$d=5.1$ kpc. We can compare our results with those obtained using the number
of ionizing photons given by \cite{mart05} for main sequence stars, in this
case the distance is 8.8 kpc for an O3V and 7.4 kpc for an O4V.

We cannot take these results at their face value, given the limitations of the
method described above. However, the results can be
used as upper limits to the distance to IRS2. The distance to W51 North, which
may be associated with IRS2, quoted in the literature were obtained through
observations of proper motions of water masers and amounts to 8.3 ($\pm$ 2.5)
kpc \citep{schn81} and 6.1 ($\pm$ 1.3) kpc \citep{imai02}. The upper limits
obtained above are compatible (within the errors) with those obtained from the
water masers if we consider the ionizing star of W51d an O3--O4 star in the
ZAMS. However, these upper limits would make IRS2 part of W51 Main just as an
effect of projection on the sky, given the distance of 2 kpc obtained with
spectroscopic parallax by \cite{fig08}.

\section{Conclusions\label{conc}}

We presented $K$-band spectra of the embedded sources IRS2E and IRS2W/W51d in
the compact HII region IRS2. The high angular resolution achieved by NIFS and
Altair made possible the separation of the strong extended emission from the
emission of the MYSO itself. It has proved a valuable tool in the study of MYSOs
as we summarize our results presented in the previous sections as follow:

\begin{itemize}

\item[1)] We identified the ionizing source of the UCHII region W51d
  as a massive star earlier than about type O4 based on the NIII
  multiplet (in emission), the HeII line (in absorption) and the
  absence of CIV in emission;

\item[2)] We detected compact (unresolved) CO overtone emission toward
  source IRS2E for the first time, suggesting the presence on an
  accretion disc around this source. The absence of radio emission and
  the presence of an accretion disk indicate that IRS2E is a massive
  YSO.

\item[3)] Combining the spectroscopy of W51d and its radio data, we found an
upper limit to the distance to IRS2 between 5.1--5.8 kpc for an ionizing star of
spectral type between O3--O4 in the ZAMS.

\end{itemize}

The fact that we detected two very massive stars, close in space, one which has
cleared its surrounding material enough to probe its photosphere and another
in an younger phase, deeply embedded in its dust cocoon and still surrounded by
a disk, indicates that the evolutive timescale of MYSOs is extremely short.

\acknowledgments

CLB and AD are grateful to FAPESP (06/02467-0) and CNPq for financial support.
PSC appreciates the continuing support from NSF. The authors wish to acknowledge
Dr. Churchwell for making available the electronic radio map of W51 IRS2 and the
anonymous referee for many valuable comments and suggestions.

Based on observations obtained at the Gemini
Observatory, which is operated by the Association of Universities for Research
in Astronomy, Inc., under a cooperative agreement with the NSF on behalf of the
Gemini partnership: the National Science Foundation (United States), the Science
and Technology Facilities Council (United Kingdom), the National Research
Council (Canada), CONICYT (Chile), the Australian Research Council (Australia),
CNPq (Brazil) and SECYT (Argentina). Partially based on observations made with
ESO Telescopes at Paranal Observatories under program ID 71.C-0344

{\it Facilities:} \facility{Gemini:Gillet}, \facility{NIFS}, \facility{VLT},
\facility{NACO}.

\clearpage

\clearpage

\clearpage

\clearpage

\end{document}